# Influence of chemical stability on the fabrication of MnGa-based devices

Lijun Zhu[1,2*] and Jianhua Zhao[1]

*1. State Key laboratory of Superlattice and microstructures, Institute of Semiconductors, Chinese Academy of Sciences*
*2. Cornell University, Ithaca, New York 14850, USA*
*lz442@cornell.edu

Ferromagnetic films of $L1_0$-ordered MnGa have shown promise not only in the applications in ultrahigh-density magnetic recording and spintronic memories, oscillators, and sensors, but also in controllable studies of novel electrical transport phenomena. However, the stability of MnGa in chemicals and oxygen plasma that are commonly used in the standard micro-/nano-fabrication process has unsettled. In this work, we report a systematic study on the chemical stability of the MnGa films in acids, acetone, ethanol, deionized water, tetramethylammonium hydroxide (TMAOH) and oxygen plasma. We find that MnGa is very stable in acetone and ethanol, while can be attacked substantially if soaked in TMAOH solution for sufficiently long time. Deionized water and acids (e.g., HCl, $H_3PO_4$ and $H_2SO_4$ solutions) attack MnGa violently and should be avoided whenever possible. In addition, oxygen plasma can passivate the MnGa surface by oxidizing the surface. These results provide important information for the fabrication and the integration of MnGa based spintronic devices.

## 1. Introduction

Ferromagnetic MnGa films with $L1_0$ long-range crystalline ordering show promise for both the spintronic technology and correlated electrical transport phenomena [1-3]. It has been established that MnGa films can have giant perpendicular magnetic anisotropy [4,5], ultrahigh coercivity [4,5], low Gilbert damping constant [6], strong magneto-optical Kerr effect [7,8], and high Curie temperature[5]. These fascinating magnetic properties make MnGa films a very compelling candidate material for ultrahigh-density perpendicular magnetic recording, nonvolatile spin-torque magnetic random access memories [9-12], terahertz spin oscillators, and linear magnetic sensors [13]. Meanwhile, the tunable magnetism and structural disorders make MnGa films an excellent playground for novel electrical transport phenomena, e.g., the topological Hall effects due to the stabilization of magnetic skyrmions [14], the orbital two-channel Kondo effects due to the coherent tunneling of two-level systems [15], and the anomalous Hall effect [16].

For the practical application in electronics and transport research, it is crucial to understand the stability of MnGa films in the micro-/nano-fabrication processes. The two key types of structural stability are thermal stability and chemical stability. $L1_0$-MnGa films are thermally stable at least at up to 800 K [17] in bulk and at 350 ºC in contact with GaAs [5], which make MnGa compatible with the CMOS technology. However, it has remained unsettled as to the chemical stability of MnGa in the commonly used chemicals and oxygen plasma during micro-/nano-fabrication. In this work, we report a systematic study of the chemical stability of MnGa films in the representative chemical solutions that are in micro-/nano-fabrication processes and oxygen plasma. Our results indicate that the chemical stability must be carefully considered during the fabrications of the MnGa-based functional devices.

## 2. Experiments and discussions

### 2.1 Growth of MnGa samples

For this study, we prepared 30 nm MnGa epitaxial films on semi-insulating GaAs (001) substrate (see Figure 1(a)) by molecular-beam epitaxy [4,5]. The base pressure was below $1 \times 10^{-9}$ mbar. A 1.5 nm MgO capping layer was used to protect the MnGa layer from oxidation in the atmosphere [5,18]. The composition was designed to be $Mn_{50}Ga_{50}$ by carefully controlling the Mn and Ga fluxes during growth and later verified by high-sensitivity x-ray photoelectron spectroscopy (XPS) measurements (Thermo Scientific ESCALAB 250Xi) with Al $K\alpha$ source and relative atomic sensitivity factors of 13.91 (1.085) for Mn $2p$ (Ga $3d$). As shown in Figure 1(b), MnGa films show asymmetric elemental XPS peaks for both Mn $2p$ and Ga $3d$ spectrums. The asymmetry of these peaks should be attributed to the significant shielding effect to the core levels by the high-density at the Fermi level due to the alloy nature of the film, which further reveals the good protection of a 1.5 nm MgO layer from oxidation.

### 2.2 Chemical stability in acid solutions

We first exam the chemical stability of MnGa flowing the workflow in Figure 2(a) in acid solutions that are popular in the wet etching process. A 30 nm MnGa film protected by a 1.5 nm MgO layer was first annealed at 110 ºC in vacuum for 20 minutes and then coated with an ultrathin HMDS layer and a 1 μm photoresist of AZ6130. The patterns were reasonably transferred from the photomask onto the sample surface after the ultraviolet exposure with a Suss MicroTec MA/BA6 Contact Mask Aligner, the development in tetramethylammonium Hydroxide (TMAOH) solution (TMAOH:$H_2O$=4:1), a 10 second rinsing in deionized water, drying with nitrogen gas flow, and the removal of the residue resist in oxygen plasma. As a reference, we first performed dry etching with argon ion milling. As indicated by the optical microscopy imaging in Figure 2(b), the dry etching yields reasonably defined patterns with sharp smooth edges after cleaning off the photoresist with acetone and ethanol. This indicates that the argon-ion milling is a highly selective etching process, with the etching rate much faster in the film normal direction than the in-plane direction. However, the wet etching using acid solutions, i.e., HCl, $H_3PO_4$ and $H_2SO_4$ solutions, result in strong sidewall etching, regardless of the combination of different volume ratio and etching time. As shown in Figure 2(c), all the edges are very rough, and some patterns which were supposed to stay was gone even before the parts which were intended to be etched away still stay on the sample. We also note that some of the connection wires between big pads, which are 10 or 20 μm wide and covered with the resist, were completely etched away even before some of the regions that were not covered with the resist. This observation clearly indicates that the wet etching of MnGa is rather non-uniform with very quick sidewall etching and that the resist seems to be unable to completely protect the MnGa film beneath it, especially in the edge regions. We conclude that acid solutions should be avoided in the fabrication processes of MnGa devices whenever possible.

### 2.3 Chemical stability in non-acid solutions and oxygen plasma

Now we consider the robustness of the MnGa films against the organic solutions, bases, deionized water, and oxygen plasma that are very popular in the standard semiconductor fabrication process. Acetone and ethanol are the most commonly used organic solutions for sample cleaning and lift-off procedures. A typical base for developing the ultraviolet exposed photoresists, such as positive-tone AZ6130 and negative-tone L300, is the TMAOH solution. Deionized water is widely used to rinse samples after photoresist development and to dilute acid or base solutions. Oxygen plasma is usually introduced to etch the resist and to clean the residual resists after photolithography development. Here we use the resistance ($R$) of the MnGa layer as a measure of the degree of the attack of the MnGa samples in various types of chemical environment (see Figure 3(a) for a schematic of the measurement geometry). In Figure 3(b) we plot the resistance of the 3 mm-wide and 6 mm-long MnGa film pieces with a 1.5 nm MgO capping layer as a function of soaking time in different chemicals. $R$ remains constant in the acetone and the ethanol solutions, indicating that MnGa is not attacked at all by the acetone and ethanol. In striking contrast, in deionized water $R$ increases dramatically with increasing soaking time and reaches 20 MΩ in 5 minutes, revealing a rather violent reaction between MnGa and $H_2O$. Therefore, the use of chemicals containing water should be avoided or reduced during the device fabrication of the MnGa whenever possible. For example, cleaning of the developer off the sample after photolithography development should be done via a quick rinsing rather than a long time soaking in water. In the 75% TMAH developer, the resistance change of the MnGa sample is negligible in the first 3 minutes, while the extended soaking can increase the resistance by a factor of 10 due to a relatively slow attack. This indicates a short development time, e.g., < 1 min, is not a concern for the MnGa devices. Oxygen plasma with the power of 200 W can oxidize the MnGa surface and passivate the surface in the first 1 min, which indicates that influence of the oxygen plasma can be substantial for an ultrathin MnGa film while negligible for relatively thick films.

### 3. Conclusion

We have systematically studied the chemical stability of the MnGa films in acids (e.g., HCl, $H_3PO_4$ and $H_2SO_4$), acetone, ethanol, deionized water, TMAOH developer, and oxygen plasma. We find that MnGa is very stable in acetone and ethanol, while can be attacked substantially if soaked in TMAOH solution for sufficiently long time. Deionized water and acids attack MnGa violently and should be avoided whenever possible. In addition, oxygen plasma can passivate the MnGa surface by oxidizing the surface. These results provide important information for the fabrication and the integration of MnGa based spintronics devices.

## Acknowledgments

The work was supported by the National Program on Key Basic Research Project [MOST, Grant Nos. 2018YFB0407601, 2015CB921500], Key Research Project of Frontier Science of Chinese Academy of Science [Grant Nos. QYZDY-SSW-JSC015, XDPB12].

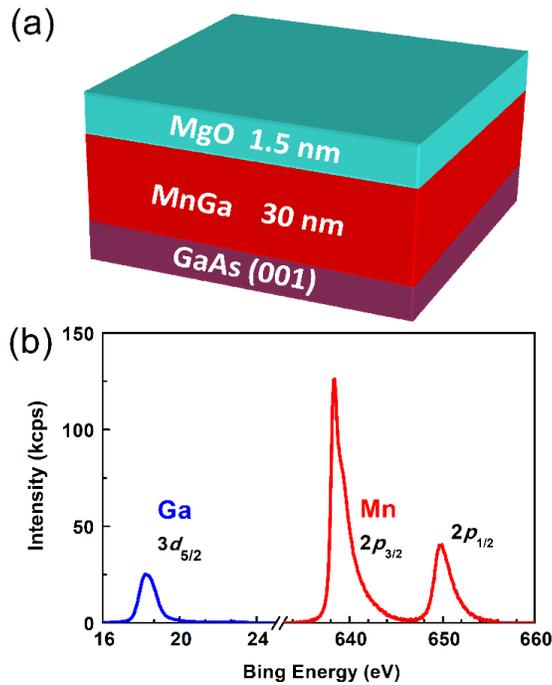

Figure 1. (a) Schematic of the sample structure. (b) XPS patterns of the MnGa layer, indicating a Mn: Ga atomic ratio of 1:1.

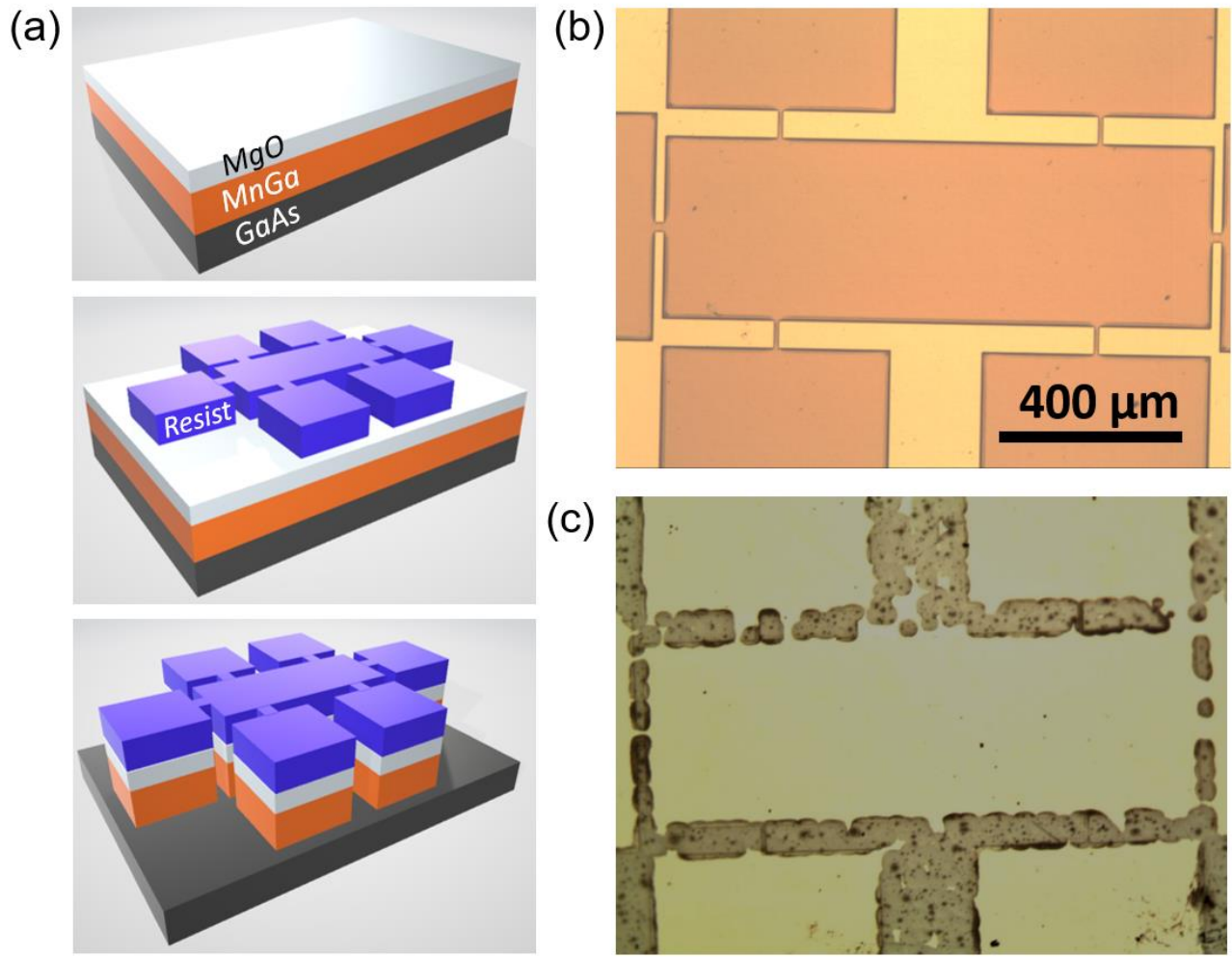

Figure 2. (a) Schematic depiction of the etching test workflow: the thin film before photolithography (top), resist patterns after photolithography (middle) and patterns after etching (bottom). Optical microscopy images for the MnGa sample surfaces patterned by (b) dry etching and (c) wet etching.

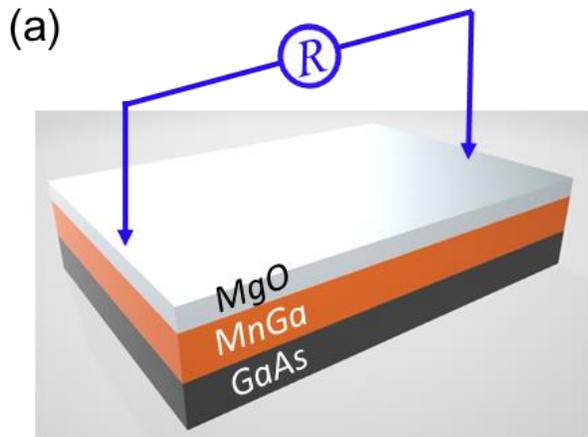

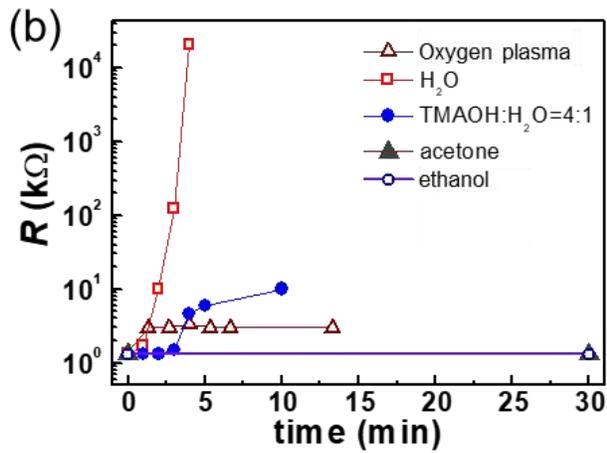

Figure 3. (a) Schematic of the resistance measurement geometry. (b) Evolution of the resistance of a 30 nm MnGa sample in ethanol, acetone, TMAOH developer, deionized water, and oxygen plasma as a function of time.